\documentclass{article}
\usepackage{graphicx}
\usepackage{subcaption}
\usepackage{authblk}
\usepackage{url}
\usepackage{booktabs}
\usepackage{amssymb}
\usepackage[margin=1in]{geometry}
\title{IndoorGNN: A Graph Neural Network based approach for Indoor Localization using WiFi RSSI}
\author{
  Rahul Vishwakarma, Rucha Bhalchandra Joshi, and
  Subhankar Mishra \\
  \affil[1]{}{National Institute of Science Education and Research, Bhubaneswar\\
  Homi Bhabha National Institute, Mumbai, India} \\
\vspace{0.2cm}
\texttt{\{rahul.vishwakarma,rucha.joshi,smishra\}@niser.ac.in}}

\date{} 

\begin{document}
\maketitle

%
%

%
%
\begin{abstract}
Indoor localization is the process of determining the location of a person or object inside a building. Potential usage of indoor localization includes navigation, personalization, safety and security, and asset tracking. Some of the commonly used technologies for indoor localization include WiFi, Bluetooth, RFID, and Ultra-wideband. Out of these, WiFi's  Received Signal Strength Indicator (RSSI)-based localization is preferred because WiFi Access Points (APs) are widely available and do not require additional infrastructure or hardware to be installed.

We have two main contributions. First, we develop our method, 'IndoorGNN'. This method involves using a Graph Neural Network (GNN) based algorithm in a supervised manner to classify a specific location into a particular region based on the RSSI values collected at that location.

Most of the ML algorithms that perform this classification require a large number of labeled data points (RSSI vectors with location information). Collecting such data points is a labor-intensive and time-consuming task. To overcome this challenge, as our second contribution, we demonstrate the performance of IndoorGNN on the restricted dataset. It shows a comparable prediction accuracy to that of the complete dataset.
We performed experiments on the UJIIndoorLoc and MNAV datasets, which are real-world standard indoor localization datasets. Our experiments show that IndoorGNN gives better location prediction accuracies when compared with state-of-the-art existing conventional as well as GNN-based methods for this same task. It continues to outperform these algorithms even with restricted datasets. It is noteworthy that its performance does not decrease a lot with a decrease in the number of available data points.

Our method can be utilized for navigation and wayfinding in complex indoor environments, asset tracking and building management, enhancing mobile applications with location-based services, and improving safety and security during emergencies.
Our code is available at: \url{https://gitlab.niser.ac.in/smlab-niser/23indoorgnn}

\end{abstract}

\textbf{Keywords:} Indoor Localization, Graph Neural Network, WiFi, Received Signal Strength Indicator

\section{Introduction}

The process of determining the location of a person or a device in a given datum is called localization. 
When localization is done within an indoor environment, it is called Indoor Localization. 
Potential usage of indoor localization includes navigation\cite{navigation,navigation1}, asset tracking\cite{asset_trace}, safety and security\cite{emerg_evac}, personalization\cite{analytics&person}, healthcare \cite{helthcare,helthcare1}, and analytics\cite{analytics&person}. 
For outdoor localization, satellite-based navigation system like GPS (Global Positioning System) is commonly used because of their good performance. But when it comes to indoor localization, GPS doesn't perform well as indoor environments consist of obstacles like walls, floors, ceilings, and other materials that can block the satellite signals\cite{gps_indoor}. Also, the GPS receiver needs to receive signals from at least four satellites to find the location accurately, which becomes harder in indoor environments.

One example of an indoor environment where GPS fails is a large shopping mall or department store. These indoor spaces typically have multiple floors, complex layouts, and dense structures such as walls and ceilings. GPS signals from satellites struggle to penetrate through these obstacles, leading to weak or no signal reception indoors. As a result, traditional GPS-based positioning methods become unreliable or completely ineffective. In such cases, an alternative indoor positioning system is necessary to provide accurate location information.

\begin{figure}
  \centering
  \includegraphics[width=1\textwidth]{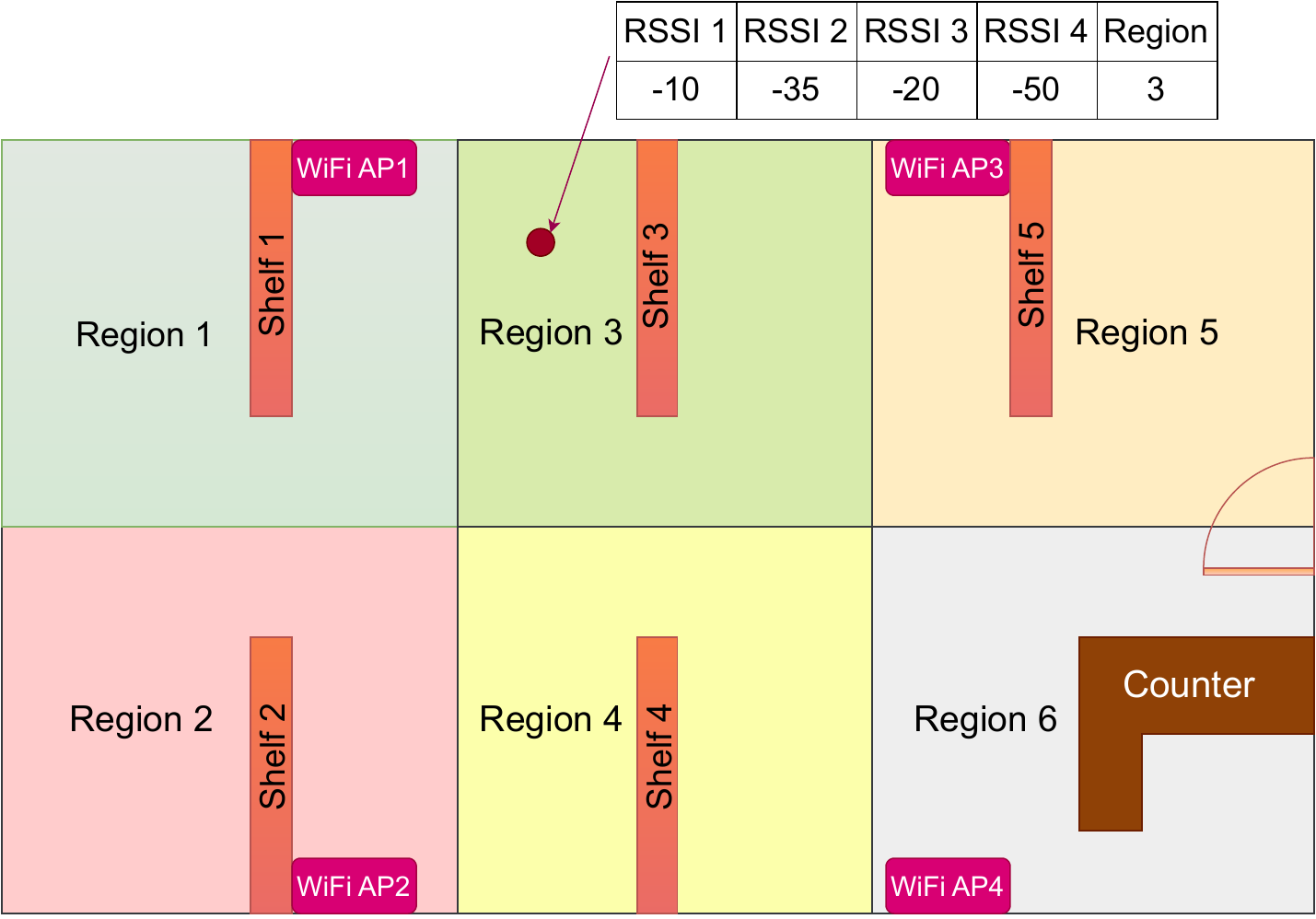}
  \caption{Illustratration of the regions within a shopping mall and the RSSI vector at a specific point}
  \label{fig:mall}
\end{figure}

To tackle the problem of indoor localization, various technologies like Bluetooth\cite{bluetooth_indoor_1,bluetooth_indoor_2,bluetooth_indoor_3}, RFID\cite{rfid_1,rfid_2}, Ultra-wideband\cite{uwb_1,uwb_2}, and WiFi\cite{wifi_1,wifi_2} are explored. Out of these technologies, WiFi RSSI-based indoor localization is preferred\cite{wifi_best_survey} because WiFi access points are generally available in most buildings, and smartphone has WiFi modules that can be used as a receiver, so it does not require any additional infrastructure or hardware to be installed. 
However, indoor localization using WiFi RSSI fingerprint requires a large amount of labeled data to train a Machine Learning model\cite{large_labelled_data}. Getting the labeled data which consists of an RSSI vector at a point and the location information of that point, is a labor-intensive and time-consuming task.

Figure \ref{fig:mall} depicts the creation of various regions inside shopping malls for indoor localization, enabling a range of applications. These include targeted advertising and promotions, where retailers can deliver personalized advertisements and offers based on shoppers' proximity to specific stores or product categories. Location-based services enhance the shopping experience by providing notifications about nearby deals, discounts, or events. Indoor localization also facilitates crowd management, allowing mall operators to monitor and manage crowd flow and congestion in real-time. Additionally, the use of indoor localization data provides valuable insights into shopper behavior, footfall patterns, and popular areas within the mall, enabling informed decisions on store placement, product positioning, and overall mall design to optimize customer satisfaction and revenue generation.

In this paper, we present an approach for label/region prediction by leveraging the power of Graph Neural Networks (GNN). Our proposed model, named IndoorGNN, utilizes a GNN-based methodology that constructs a graph representation where each node corresponds to a specific RSSI vector. The edge weights in the graph are determined by the similarity between the RSSI vectors. To evaluate the effectiveness of our approach, we conducted experiments using two widely recognized datasets: UJIIndoorLoc\cite{ujiindoorloc} and MNAV\cite{museum}. These datasets are extensively employed in the domain of indoor localization and serve as standard benchmarks for comparing the performance of various models. The experimental results demonstrate that IndoorGNN outperforms state-of-the-art algorithms such as Graph Neural Network (GNN) based models where APs are considered nodes\cite{indoor_gnn}, k-Nearest Neighbor (kNN), Support Vector Machine (SVM), and Multilayer Perceptron (MLP) in terms of classification accuracy. Moreover, even when trained on a reduced fraction of the dataset, IndoorGNN consistently maintains its superior performance over other algorithms.

The contributions of our research can be summarized as follows:
\begin{itemize}

\item Development and detailed model description: We introduce IndoorGNN, a novel model for accurate label prediction in indoor localization. We provide a comprehensive description of its architecture and key components, enabling a deeper understanding and facilitating its implementation in future research and practical applications.

\item Evaluation of IndoorGNN: We conduct a thorough evaluation of the performance of IndoorGNN in accurate label prediction tasks for indoor localization. This assessment allows us to measure the effectiveness and efficiency of the model in solving real-world problems.

\item Effectiveness of IndoorGNN: Our experimental results indicate the better performance of IndoorGNN in comparison to existing algorithms. We observe this effectiveness across both complete and partial training datasets, highlighting the robustness and versatility of IndoorGNN across different scenarios and dataset sizes.

\end{itemize}

Our paper is structured as follows:
Section \ref{sec:rel_works} provides an overview of the related works that have been conducted to enhance ML-based indoor localization systems. We review the existing literature and highlight the advancements made in this field. 
In Section \ref{sec:prob_stat}, we present a mathematical definition of the problem and formulate it as a Graph Neural Network (GNN) based model. We outline the key components and outline how GNN can effectively address the challenges associated with indoor localization. 
Section \ref{sec:method} details the methodology behind our GNN-based model for indoor localization. We provide a step-by-step explanation of our approach, including the construction of the graph representation with RSSI vectors as nodes and the utilization of edge weights derived from vector similarity. 
Moving forward, in Section \ref{sec:expt}, we describe the experimental setup employed in this study. We discuss the dataset used for evaluation and provide a thorough explanation of how we approached solving the problems identified in Section \ref{sec:prob_stat}. This section highlights the experimental methodology with the results and enables readers to understand the practical aspects of our research.

\section{Related works}
\label{sec:rel_works}
The propagation of an electromagnetic signal adheres to the inverse square law, but the presence of obstacles obstructs the signal path, making it challenging to formulate the decrease in signal strength as an inverse square law. To address this challenge, various Machine Learning algorithms have been explored to model signal decay and generate RSSI fingerprints for predicting the location of new RSSI vectors. Commonly used algorithms for indoor positioning include K-Nearest Neighbors (KNN) \cite{knn_1,knn_2}, Decision Trees \cite{dt_1}, Support Vector Machines (SVMs) \cite{svm_1,svm_2}, and Artificial Neural Networks (ANNs) \cite{dl_1,dl_2,dl_3,dl_4}. Among these, kNN, SVM, and Neural Network models have exhibited superior performance \cite{kNN_better}. However, these models do not consider the geometric characteristics of the indoor localization problem.

In recent years, Graph Neural Network (GNN) based models have emerged as a promising approach for indoor localization. In the paper \cite{dom_adv}, a novel approach utilizing heterogeneous graphs based on the RSSI between Access Points (APs) and waypoints was proposed to comprehensively represent the topological structure of the data. Another paper \cite{a_geom} discussed the limitations of traditional ML algorithms in effectively encoding fingerprint data and proposed a novel localization model based on a GraphSAGE estimator. Moreover, \cite{indoor_gnn} introduced a GNN-based model that leverages the geometry of the Access Points for WiFi RSSI-based indoor localization. Further in \cite{gnn_indoor} authors handle the problem of indoor localization with GNN where they consider the APs as nodes and edges between them are decided by the received power between them and they perform the experiments with both homogeneous and heterogeneous graphs.

However, one of the inherent challenges in indoor localization is the fluctuation of signal strength in WiFi RSSI data. The varying signal strengths make it difficult to accurately determine the location of a user or object in an indoor environment.
To mitigate the impact of signal strength fluctuations, we introduce a graph-based fingerprinting method for indoor localization. Our approach focuses on the stability of proximity patterns in RSSI data rather than relying solely on individual AP RSSI values. By leveraging these stable patterns, our method enhances the accuracy and reliability of indoor localization, providing a robust solution for overcoming the challenges associated with fluctuating signal strengths in indoor environments.


\section{Problem Statement}
\label{sec:prob_stat}
\subsection{Preliminaries}
A graph $\mathbf{G=(V, E)}$ consists of a set of vertices $\mathbf{V}$ and a set of edges $\mathbf{E}$. 
The edge set is given as $\mathbf{E} = \{(i, j) \, | \, i, j\in \mathbf{V}$ \}. The feature vectors corresponding to edge $(i, j)$ is represented as $\mathbf{e}_{ij}$. The neighborhood of a node $i$ is represented as $\mathcal{N}(i)$, such that $\mathcal{N}(i) = \{j\, |\, (i, j) \in \mathbf{E}\}$.
A node $i$'s representation in $l^{th}$ GNN layer is given by $\mathbf{x}_i^{(l)}$.

\subsection{Problem Definition}
Let $ \mathbf{F} \in \mathbb{R} ^ {n \times m} $ be the feature matrix with $ n $ data points, each with an RSSI vector of size $ m $, where $ \mathbf{F}_{i, j} $ represents the signal strength of $ j $ th access point in the $ i $th data point.
Let $\mathbf{L} \in \mathbf{T}^n$ be the matrix of labels for the prediction task, where $\mathbf{L}_i$ represents the region of the $i$th data point and $\mathbf{T}$ be the set of regions for classification. 
We define a training mask $\mathbf{M}$, for the data points in the training set with $\mathbf{M} \in \{0,1\}^{n}$ which represents the availability of data point for training, where $\mathbf{F}_{i}$ is available for training only if $\mathbf{M}_{i} = 1$. 
Let us define train ratio $\mathbf{r}$ as $\mathbf{r} = \frac{1}{n}\sum_i \mathbf{M_i}$ which represents the ratio of the training data points available for training the model. 
The problem that we address here is the classification of a given $\mathbf{F}_{i}$ into one of the $region \in \mathbf{T}$.

\section{Methodology}
\label{sec:method}
To categorize the positions into distinct regions using RSSI values, we employ a technique called \textit{DynamicEdgeConv}. 
The initial step involves structuring the data into a graph format using the k-Nearest Neighbors (kNN) algorithm. This approach is chosen because it effectively captures the local spatial relationships within the data, which are crucial for indoor localization. In essence, each data point (representing an RSSI measurement) is considered as a node in the graph, and edges are formed between nodes based on their proximity, determined by the kNN algorithm. This results in a graph where nodes represent data points, and edges signify the spatial relationships between them.

The DynamicEdgeConv method introduces a dynamic graph generation process within the layers of the Graph Neural Network. Rather than using a fixed, predefined graph structure, a new graph is generated at each layer of the GNN based on the kNN approach. This dynamic graph generation strategy allows the GNN to adapt and capture different levels of spatial information as it progresses through its layers. It essentially incorporates local spatial relationships into the learning process, enabling the model to effectively exploit the inherent structure and dependencies within the RSSI data. This adaptability is a critical factor in the success of the IndoorGNN model, as it ensures that the model can learn and generalize effectively in diverse indoor environments, where the spatial characteristics may vary significantly. In the subsequent section, we will provide a more in-depth and comprehensive description of the DynamicEdgeConv methodology, including its mathematical foundations and practical implementation details.

\subsection{Modeling Data as Graph}

We address the problem of indoor localization by modeling it as a graph-based problem. 
Specifically, we translate this problem to a problem where we can make use of Dynamic Edge Convolution (\textit{DynamicEdgeConv}) \cite{wang2019dynamic}.

Every data point with a feature vector containing the RSSI values of dimension $m$ collected at the particular location and the location information is treated as a point in a space $\mathbb{R}^m$. We assume that these RSSI values lead us to determine the location of the data point. 
The corresponding location information of the data points is the local region to which the data point belongs. 

A directed graph $\mathbf{G}=(\mathbf{V}, \mathbf{E})$ is constructed as a k-Nearest Neighbor graph of $\mathbf{F}$ in $\mathbb{R}^{m}$. We also include self-loops to the graph so that the point contributes to its next layer representation in a GNN. Hence every point with an RSSI value is modeled as a node in the graph $\mathbf{G}$, and the edges are constructed based on k-NN.

\subsection{Graph Neural Network for Prediction}\label{sec:gnn_pred}

The RSSI values act as the node features. The edge features are constructed using a learnable function $h_\Theta$ where $\Theta$ is the set of learnable parameters. The edge features are given as $\mathbf{e}_{ij} = \mathbf{h}_\Theta(\mathbf{x}_i, \mathbf{x}_j)$, and the learnable function $h_\Theta : \mathbb{R}^m \times \mathbb{R}^m \rightarrow \mathbb{R}^{m'}$. 

We perform edge convolution operations to compute the next layer representation of a node. The edge convolution in an intermediate operation while computing node representations \cite{joshi2021learning}. This operation is given as follows:
\begin{equation}
    \mathbf{x}'_i = \textsc{Aggregate} \left\{ \mathbf{h}_\Theta (\mathbf{x}_i, \mathbf{x}_j) j \in \mathcal{N}(i) \right\}
    \label{eq:edge_conv}
\end{equation}
where $\mathbf{x}'_i$ is the next layer representation of the node $i$, $\mathcal{N}(i) = \{ j : (i, j) \in \mathbf{E} \}$ is the set of neighbors of $i$. The aggregation function $\textsc{Aggregate}$ is RSSI value-wise symmetric. 

\begin{figure}
    \centering
    \includegraphics[width=\linewidth]{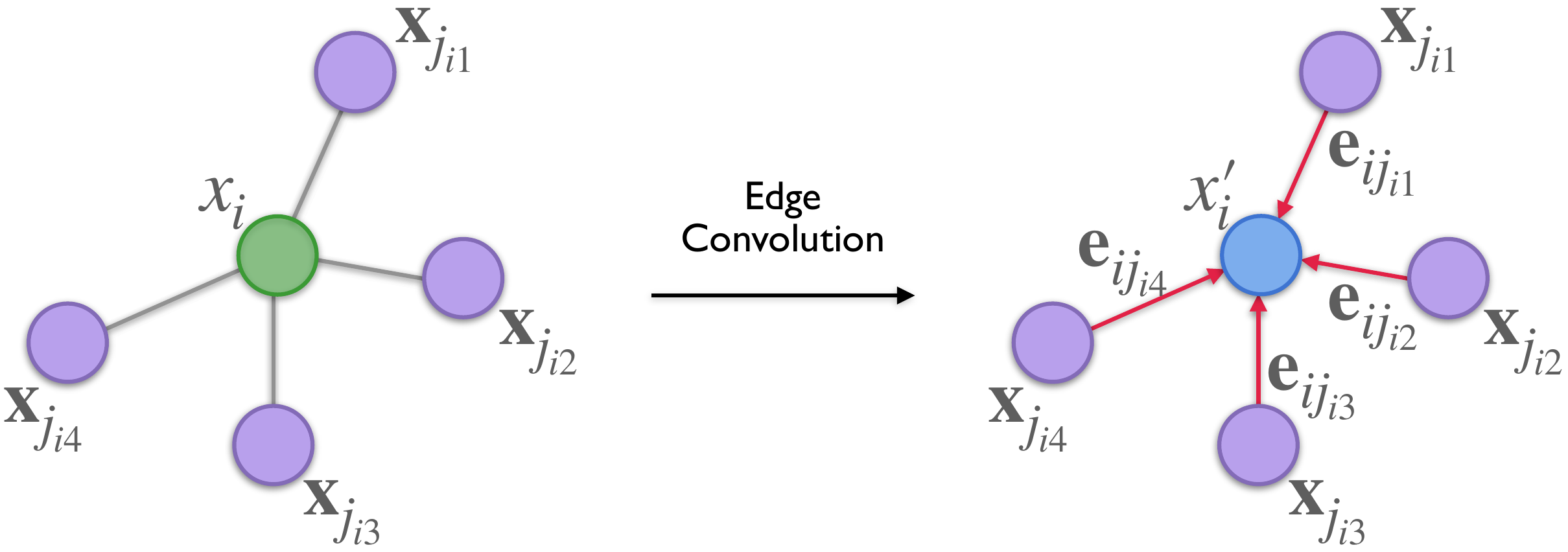}
    \caption{Edge Convolution}
    \label{fig:edge_conv}
\end{figure}

\begin{figure}
    \centering
    \includegraphics[width=0.6\linewidth]{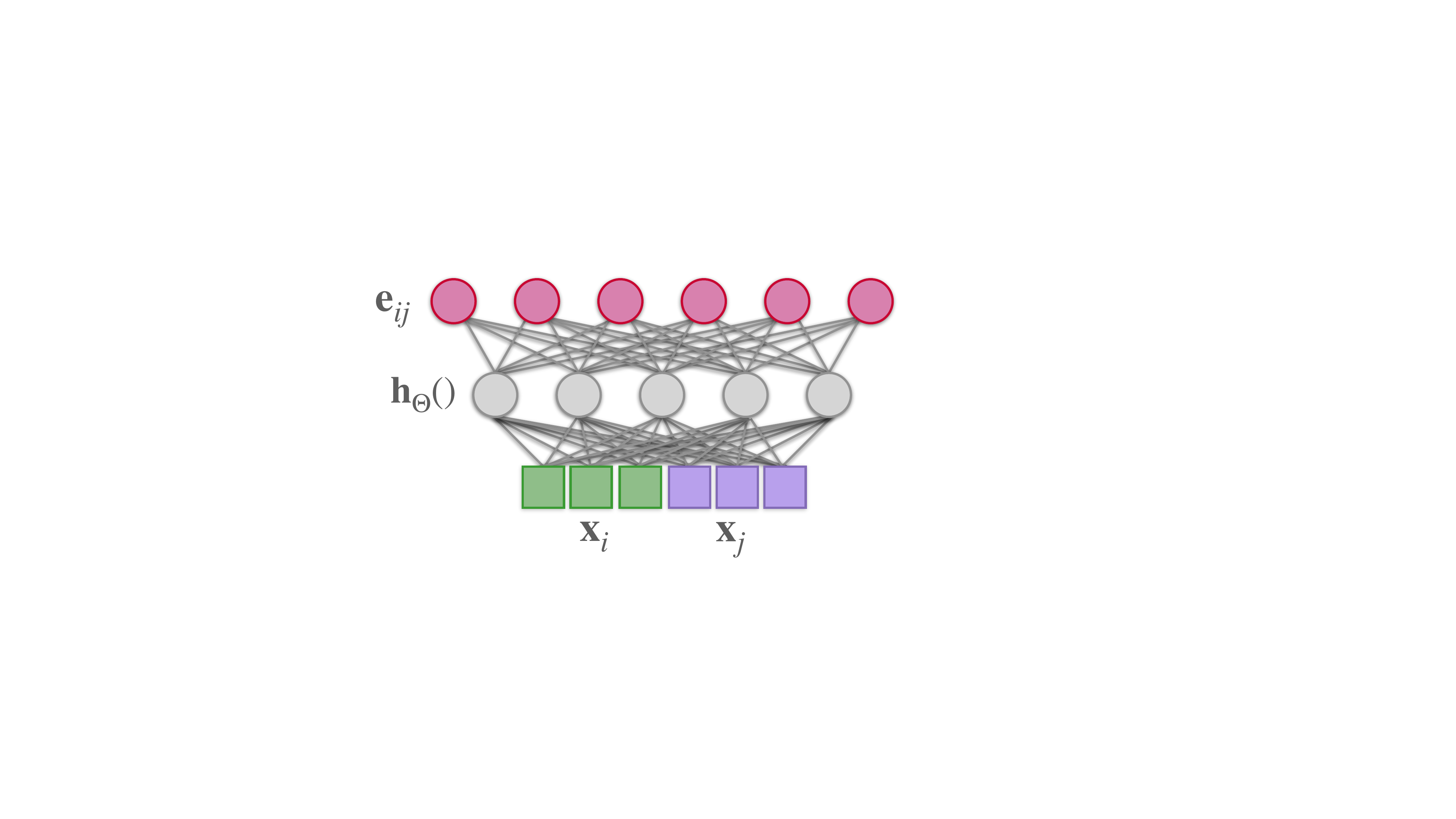}
    \caption{Computation of edge features}
    \label{fig:edge_feature_computation}
\end{figure}

Figure \ref{fig:edge_conv} shows how edge convolution takes place given the features of the node and its neighborhood. As an intermediate step, edge features are computed using the features of two nodes connected by the edges. With these edge features, a new representation is generated for the node. Figure \ref{fig:edge_feature_computation} shows in detail how the edge features are computed given the features of the two nodes connected by the edge.

In Edge Convolution, we determine the edges using k-NN. In the \textit{DynamicEdgeConvolution}, the graph is recomputed at every layer based on the node representations (features) generated at every layer of the GNN. Hence at every layer $l$, a new graph $\mathbf{G}^{(l)} = (\mathbf{V}^{(l)}, \mathbf{E}^{(l)})$. The GNN is applied on this dynamically computed graph, with the newly generated features at the particular layer.

\section{Experiments}
\label{sec:expt}
\subsection{Experimental Setup}

\subsubsection{Dataset}
To evaluate the performance of our model compared to the baseline models, we conducted experiments using the UJIIndoorLoc\cite{ujiindoorloc} and MNAV\cite{museum} datasets.

The \textbf{UJIIndoorLoc} dataset contains information about WiFi signal strengths used for testing indoor positioning systems. It covers three buildings at Universitat Jaume I, each with four or more floors and a total area of about 110 square meters. The dataset includes 19,937 training data points and 1,111 test data points, with a total of 529 attributes. These attributes provide details such as WiFi signal readings, latitude, longitude, and other helpful information. In the dataset, 520 Access Points (APs) were detected, and their signal strengths, known as RSSI (Received Signal Strength Indicator), are given in the first 520 attributes. The RSSI values range from -104dBm (weak signal) to 0dBm (strong signal), while a value of 100 means that the access point was not detected. This dataset is publicly available on the \textit{UCI Website}\footnote{\url{https://archive.ics.uci.edu/ml/datasets/ujiindoorloc}}.

The \textbf{MNAV} dataset was created for the Museo Nacional de Artes Visuales (MNAV, National Museum of Visual Arts) in Uruguay as part of the project described in paper \cite{museum}, which aimed to develop an indoor localization system utilizing WiFi fingerprinting. This dataset is publicly accessible on \textit{GitHub}\footnote{\url{https://github.com/ffedee7/posifi_mnav/}}. It comprises a total of 10,469 data points recorded during the data collection, with the detection of 188 Access Points (APs). The paper also provides a museum map that delineates the locations of the APs and defines 16 regions for the classification task. Among the 188 detected APs, 15 were installed specifically for the experiment, with each AP supporting both the 2.4GHz and 5GHz bandwidths. Consequently, the dataset encompasses 30 features derived from these APs. Here the RSSI values range from -99dBm (weak signal) to 0dBm (strong signal), while a value of 0.0 means that the access point was not detected. Table \ref{tab:dataset-summary} summarizes the details of both datasets.

\begin{table}[htbp]
\caption{Summary of Datasets}
\centering
\begin{tabular}{lrr}
\toprule
\textbf{Attributes} & \textbf{UJIIndoorLoc} & \textbf{MNAV} \\
\midrule
Total Data Points & 21,048 & 10,469 \\
Training Points & 19,937 & 8,375 \\
Test Points & 1,111 & 2,094 \\
Detected APs & 520 & 188 \\
Labels & 13 & 16 \\
RSSI Signal Range & -104dBm to 0dBm & -99dBm to 0dBm \\
Features & \hspace{0.3cm} Floor, BuildingID, Lat, & \hspace{0.3cm} RSSI with corresponding \\
 & Long, RSSI, etc. & locations \\
Dataset Availability & \textit{UCI Website} & \textit{GitHub} \\
\bottomrule
\end{tabular}
\label{tab:dataset-summary}
\end{table}

To preprocess the datasets, we made certain modifications. In both datasets, we added 104 to the RSSI feature, thereby adjusting the range to 0-104. Additionally, we replaced the missing access point values i.e. AP with an RSSI value of 100 in the UJIIndoorLoc dataset and AP with 0.0 RSSI value in the MNAV dataset, to 0. In the MNAV dataset, the set of labels (regions) $\mathbf{T}$ was pre-defined within the dataset, consisting of a total of 16 unique regions (classes). On the other hand, for the UJIIndoorLoc dataset, we generated the set $\mathbf{T}$ by combining the FLOOR and BUILDINGID columns of the dataset. This process yielded a total of 13 unique regions (classes).

\begin{figure}
  \centering
  \includegraphics[width=0.7\linewidth]{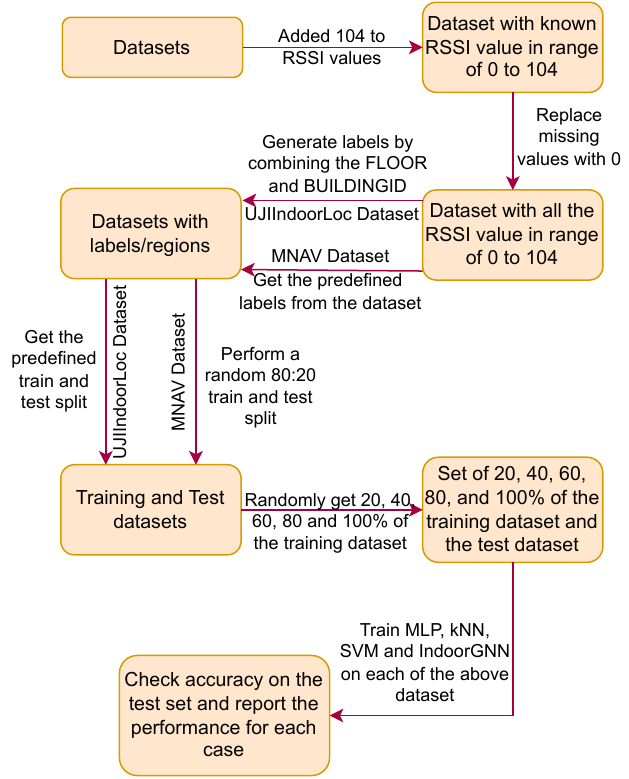}
  \caption{Flow diagram for preprocessing and training on the datasets.}
  \label{fig:flow_chart}
\end{figure}

For the UJIIndoorLoc dataset, we utilized the predefined train and test split. As for the MNAV dataset, there is no predefined train and test split, so we created a new train and test split with an 80:20 ratio. This resulted in 8,375 training data points and 2,094 test data points out of the total 10,469 data points available. We have used this training and test set for all of our experiments, and all the accuracy results presented in this paper are on the test set. Our goal here is to improve the accuracy of label (region) prediction. As getting labeled datasets is a time-consuming and labor-intensive task so to study the effect of using a restricted dataset for training, we have performed the experiments for train ratio $ \mathbf{r} \in \{0.2, 0.4, 0.6, 0.8, 1.0\}$. The flowchart in Figure \ref{fig:flow_chart} visually summarizes the sequence of tasks involved in our study.

\subsubsection{Models Description}

In the case of \textbf{MLP}, we trained a neural network on the given training data set and then assessed the performance on the test set, and the outcome derived from this was reported.

For \textbf{kNN}, a grid search approach was adopted to ascertain the ideal value for \textit{k}. Following this determination, an evaluation of its performance was conducted using the test set. Notably, we observed that as the size of the training data reduced, the optimal \textit{k} value also decreased for the task.

In \textbf{SVM}, we utilized the Radial Basis Function (RBF) as the kernel function, which exhibited superior performance compared to other kernel functions such as linear. Additionally, we conducted a grid search to identify the optimal value for the parameter \textit{C}, which balances the misclassification of training examples against the simplicity of the decision surface.

For our GNN-based baselines, we analyze \textbf{GNN}\cite{indoor_gnn}.
This paper employs Graph Neural Networks (GNNs) with APs as nodes and the distance between them as edges of the graph for indoor localization and encompasses experimental evaluations on the UJIIndoorLoc and MNAV datasets. Consequently, we use the authors' code to run on our preprocessed datasets.

\begin{figure}
    \centering
    \includegraphics[width=0.8\linewidth]{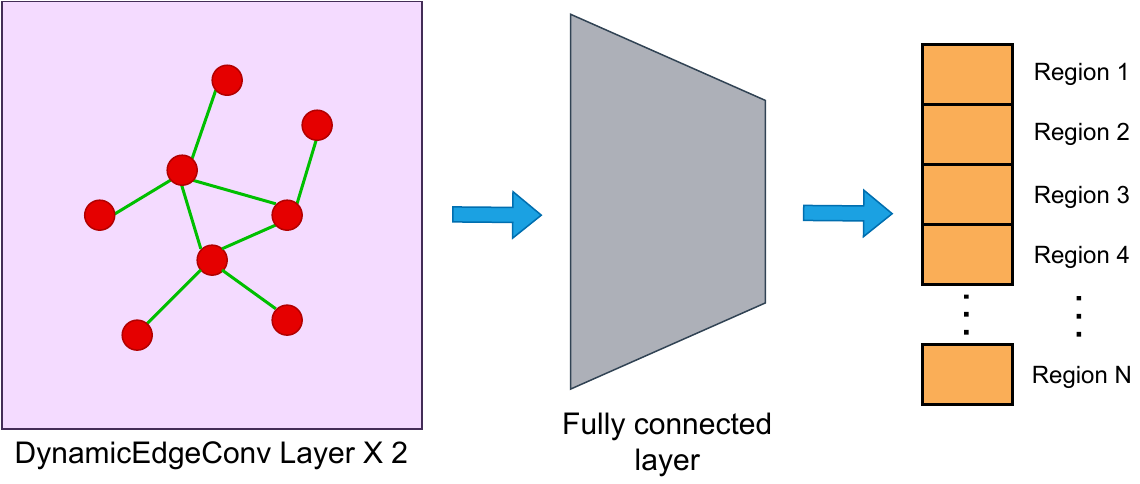}
    \caption{Architecture of IndoorGNN model}
    \label{fig:arch}
\end{figure}

In \textbf{IndoorGNN}, we employed two graph neural network layers as discussed in section \ref{sec:gnn_pred} followed by a fully connected neural network. The complete architecture of the model is shown in figure \ref{fig:arch}. In the GNN layers, we utilized the \textit{mean} function as the aggregator function in equation \ref{eq:edge_conv} when doing Edge Convolution to aggregate values from the neighboring nodes, as it demonstrated better performance compared to other aggregator functions such as \textit{max} and \textit{sum}. 

\subsection{Experimetal Results}

Table \ref{tab:complete} presents a comprehensive overview of algorithm performance on the complete datasets, including their classification accuracy. Notably, our proposed IndoorGNN algorithm consistently outperforms both conventional and GNN-based algorithms.

For the UJIIndoorLoc dataset, the preceding GNN model achieved a maximum classification accuracy of 91.3\%. Among all other algorithms, SVM demonstrated the best performance with an accuracy of 94.5\%. In contrast, our IndoorGNN algorithm stood out with the highest classification accuracy of 95.8\%.

Similarly, when considering the MNAV dataset, the GNN model achieved a classification accuracy of 96.3\%. Among the other algorithms, kNN exhibited the best performance with an accuracy of 97.5\%. Remarkably, our IndoorGNN algorithm also achieved a classification accuracy of 97.5\%, surpassing the performance of the GNN model and equalling that of the top-performing model, kNN.

These results underscore the effectiveness and competitiveness of the IndoorGNN algorithm in accurately classifying indoor locations for both datasets.

\begin{table}[h]
  \centering
  \caption{Performance matrix of different Algorithms on complete UJIIndoorLoc and MNAV datasets}
  \label{tab:complete}
  \begin{tabular}{|c|c|c|c|}
    \hline
     &  & \multicolumn{2}{|c|}{\textbf{Classifiation Accuracy}} \\
     \cline{3-4}
    \textbf{S. No.} & \textbf{Algorithm} & UJIIndoorLoc & MNAV \\
    \hline
    1 & MLP & 92.0 & 94.8 \\
    \hline
    2 & kNN & 92.5 & \textbf{97.5} \\
    \hline
    3 & SVM & 94.5 & 95.0 \\
    \hline
    4 & GNN \cite{indoor_gnn} & 91.3 & 96.3 \\
    \hline
    5 & IndoorGNN & \textbf{95.8} & \textbf{97.5} \\
    \hline
  \end{tabular}
\end{table}

\begin{figure}
     \centering
     \begin{subfigure}[b]{1\textwidth}
         \centering
         \includegraphics[width=\textwidth]{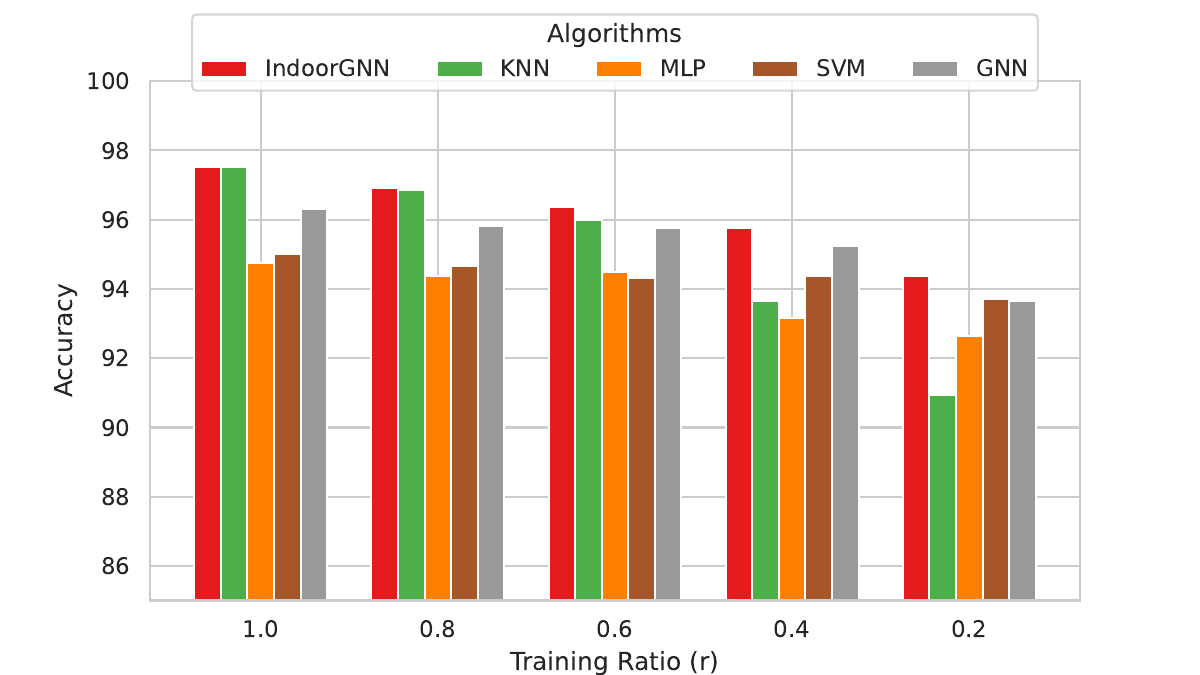}
         \caption{MNAV Dataset}
         \label{fig:manv_plot}
     \end{subfigure}
     \hfill
     \begin{subfigure}[b]{1\textwidth}
         \centering
         \includegraphics[width=\textwidth]{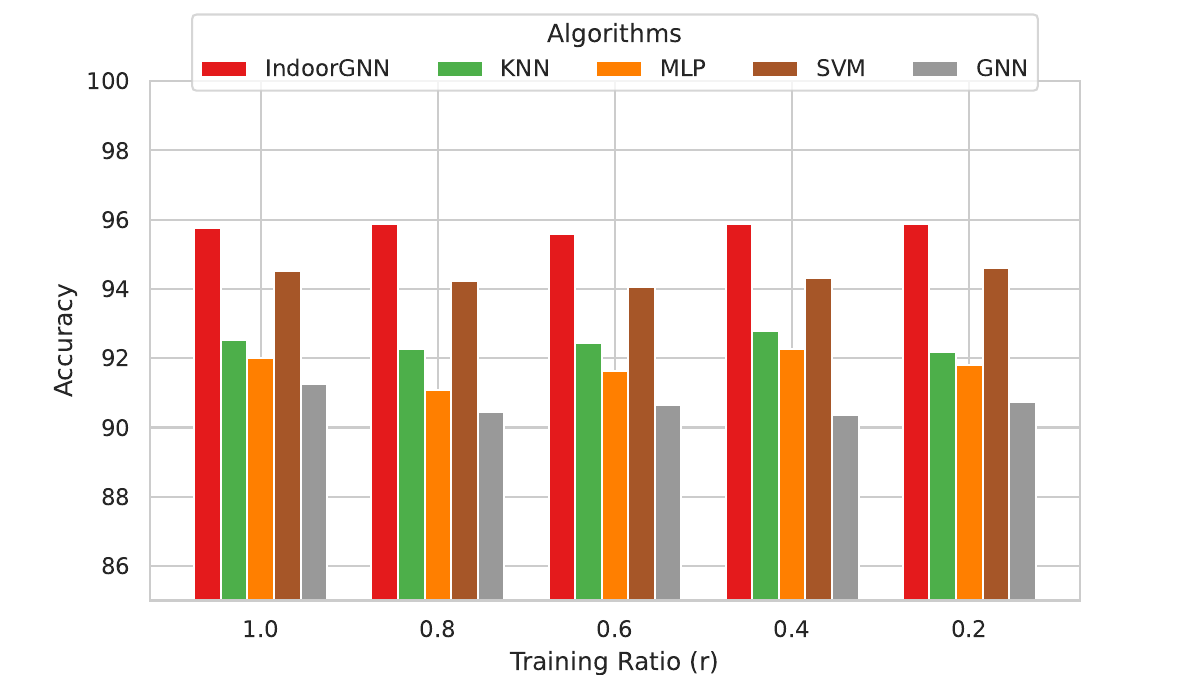}
         \caption{UJIIndoorLoc dataset}
         \label{fig:uji_plot}
     \end{subfigure}
        \caption{Comparison of Classification Accuracy for different Train Ratio on the MNAV and UJIIndoorLoc datasets.}
        \label{fig:combined_plot}
\end{figure}

Figure \ref{fig:uji_plot} presents a comprehensive assessment of algorithm performance on the UJIIndoorLoc dataset, considering varying training dataset sizes. Notably, IndoorGNN consistently outperforms all other algorithms, maintaining its superior performance even with reduced training data. Likewise, Figure \ref{fig:manv_plot} showcases the performance analysis of diverse algorithms on the MNAV dataset, across different training dataset sizes. Evidently, IndoorGNN either matches or surpasses the performance of other algorithms when trained on a fraction of the dataset.

Figure \ref{fig:combined_plot} further emphasizes these observations. For the MNAV dataset, besides IndoorGNN, kNN and GNN also exhibit strong performance. Conversely, on the UJIIndoorLoc dataset, SVM demonstrates superior performance, second only to IndoorGNN. It's worth noting that IndoorGNN consistently outperforms all others on both datasets, highlighting its ability to generalize across diverse data scenarios.

Analyzing the results depicted in Figure \ref{fig:combined_plot}, an intriguing pattern emerges regarding algorithm performance on the UJIIndoorLoc and MNAV datasets. Specifically, for the UJIIndoorLoc dataset, even as the training data points decrease, algorithm performance remains relatively stable. This phenomenon can be attributed to the presence of fixed points within the UJIIndoorLoc dataset, where multiple RSSI data measurements have been recorded. Consequently, the algorithms maintain their performance because removing a specific data point at a given location doesn't result in significant information loss, given the availability of other data points representing the same RSSI vector at that location.

In contrast, the MNAV dataset consists primarily of unique locations within the sampled area. Consequently, algorithm performance is more reliant on the complete dataset. As the number of training data points decreases, algorithm performance also diminishes accordingly.

In summary, these findings underscore the significant impact of dataset characteristics on algorithm performance. The presence of multiple measurements at fixed points in the UJIIndoorLoc dataset allows for robust performance even with reduced training data. In contrast, the unique location representation of most data points in the MNAV dataset necessitates a larger training dataset for optimal algorithm performance, with performance decreasing as training data diminishes.

\section{Conclusion}
\label{sec:concl}
In conclusion, this research introduces IndoorGNN, a cutting-edge Graph Neural Network model designed for indoor localization. Through extensive experimentation on two diverse datasets, IndoorGNN consistently demonstrates its superior performance in both complete and constrained training scenarios.

On the UJIIndoorLoc dataset, IndoorGNN shines as it outperforms all other algorithms, including GNN-based counterparts. It achieves a notable boost in accuracy, elevating it from 94.5\% to an impressive 95.8\% across the complete training dataset. This superiority extends even when dealing with partial training data, making IndoorGNN the preferred choice for resource-intensive and labor-intensive indoor localization tasks. On the MNAV dataset, IndoorGNN exhibits a marked improvement over the GNN-based model, matching the classification accuracy of the top-performing kNN algorithm at a remarkable 97.5\% across the entire dataset. Furthermore, it surpasses all alternative algorithms when applied to a limited dataset. These findings firmly establish IndoorGNN's effectiveness across a spectrum of scenarios.

In summary, IndoorGNN emerges as a compelling solution for indoor localization, consistently delivering superior performance even with smaller training datasets. Its capacity to outshine competing algorithms in challenging scenarios underscores its potential for real-world applications that demand both efficiency and precision.

Looking ahead, our future research will explore several exciting avenues:
\begin{itemize}
    \item Investigating methods for precise latitude and longitude location determination.
    \item Tackling scenarios with limited labeled data but an abundance of unlabeled data.
    \item Developing strategies to effectively handle missing RSSI values.
    \item Investigating localization methods in private settings using Graph Neural Networks \cite{joshi_dtrap}.
\end{itemize}

These endeavors will further enhance the capabilities of IndoorGNN and extend its applicability in the field of indoor localization.

\bibliographystyle{splncs04}
\bibliography{ref}

\begin{thebibliography}{10}
\providecommand{\url}[1]{\texttt{#1}}
\providecommand{\urlprefix}{URL }
\providecommand{\doi}[1]{https://doi.org/#1}

\bibitem{dl_2}
Adege, A.B., Lin, H.P., Tarekegn, G.B., Jeng, S.S.: Applying deep neural network (dnn) for robust indoor localization in multi-building environment. Applied Sciences  \textbf{8}(7), ~1062 (2018)

\bibitem{wifi_best_survey}
Al~Nuaimi, K., Kamel, H.: A survey of indoor positioning systems and algorithms. In: 2011 international conference on innovations in information technology. pp. 185--190. IEEE (2011)

\bibitem{bluetooth_indoor_2}
Bekkelien, A., Deriaz, M., Marchand-Maillet, S.: Bluetooth indoor positioning. Master's thesis, University of Geneva  (2012)

\bibitem{museum}
Bracco, A., Grunwald, F., Navcevich, A., Capdehourat, G., Larroca, F.: Museum accessibility through wi-fi indoor positioning (2020)

\bibitem{svm_2}
Chriki, A., Touati, H., Snoussi, H.: Svm-based indoor localization in wireless sensor networks. In: 2017 13th international wireless communications and mobile computing conference (IWCMC). pp. 1144--1149. IEEE (2017)

\bibitem{uwb_1}
Dabove, P., Di~Pietra, V., Piras, M., Jabbar, A.A., Kazim, S.A.: Indoor positioning using ultra-wide band (uwb) technologies: Positioning accuracies and sensors' performances. In: 2018 IEEE/ION Position, Location and Navigation Symposium (PLANS). pp. 175--184. IEEE (2018)

\bibitem{gps_indoor}
Dedes, G., Dempster, A.G.: Indoor gps positioning-challenges and opportunities. In: VTC-2005-Fall. 2005 IEEE 62nd Vehicular Technology Conference, 2005. vol.~1, pp. 412--415. Citeseer (2005)

\bibitem{emerg_evac}
Eksen, K., Serif, T., Ghinea, G., Gr{\o}nli, T.M.: Inloc: Location-aware emergency evacuation assistant. In: 2016 IEEE International Conference on Computer and Information Technology (CIT). pp. 50--56. IEEE (2016)

\bibitem{navigation1}
El-Sheimy, N., Li, Y.: Indoor navigation: State of the art and future trends. Satellite Navigation  \textbf{2}(1),  1--23 (2021)

\bibitem{helthcare}
Fernandez-Llatas, C., Lizondo, A., Monton, E., Benedi, J.M., Traver, V.: Process mining methodology for health process tracking using real-time indoor location systems. Sensors  \textbf{15}(12),  29821--29840 (2015)

\bibitem{wifi_2}
He, S., Chan, S.H.G.: Wi-fi fingerprint-based indoor positioning: Recent advances and comparisons. IEEE Communications Surveys \& Tutorials  \textbf{18}(1),  466--490 (2015)

\bibitem{knn_1}
Huang, C.N., Chan, C.T.: Zigbee-based indoor location system by k-nearest neighbor algorithm with weighted rssi. Procedia Computer Science  \textbf{5},  58--65 (2011)

\bibitem{joshi2021learning}
Joshi, R.B., Mishra, S.: Learning graph representations. In: Principles of Social Networking: The New Horizon and Emerging Challenges, pp. 209--228. Springer (2021)

\bibitem{joshi_dtrap}
Joshi, R.B., Mishra, S.: Locally and structurally private graph neural networks. Digital Threats  (sep 2023). \doi{10.1145/3624485}, \url{https://doi.org/10.1145/3624485}

\bibitem{bluetooth_indoor_3}
Kalbandhe, A.A., Patil, S.C.: Indoor positioning system using bluetooth low energy. In: 2016 International Conference on Computing, Analytics and Security Trends (CAST). pp. 451--455. IEEE (2016)

\bibitem{dl_4}
Khatab, Z.E., Hajihoseini, A., Ghorashi, S.A.: A fingerprint method for indoor localization using autoencoder based deep extreme learning machine. IEEE sensors letters  \textbf{2}(1), ~1--4 (2017)

\bibitem{asset_trace}
Lee, C.K.M., Ip, C., Park, T., Chung, S.: A bluetooth location-based indoor positioning system for asset tracking in warehouse. In: 2019 IEEE International Conference on Industrial Engineering and Engineering Management (IEEM). pp. 1408--1412. IEEE (2019)

\bibitem{indoor_gnn}
Lezama, F., Gonz{\'a}lez, G.G., Larroca, F., Capdehourat, G.: Indoor localization using graph neural networks. In: 2021 IEEE URUCON. pp. 51--54. IEEE (2021)

\bibitem{gnn_indoor}
Lezama, F., Larroca, F., Capdehourat, G.: On the application of graph neural networks for indoor positioning systems. In: Machine Learning for Indoor Localization and Navigation, pp. 239--256. Springer (2023)

\bibitem{dl_3}
Liu, C., Wang, C., Luo, J.: Large-scale deep learning framework on fpga for fingerprint-based indoor localization. IEEE Access  \textbf{8},  65609--65617 (2020)

\bibitem{analytics&person}
Liu, Y., Cheng, D., Pei, T., Shu, H., Ge, X., Ma, T., Du, Y., Ou, Y., Wang, M., Xu, L.: Inferring gender and age of customers in shopping malls via indoor positioning data. Environment and Planning B: Urban Analytics and City Science  \textbf{47}(9),  1672--1689 (2020)

\bibitem{a_geom}
Luo, X., Meratnia, N.: A geometric deep learning framework for accurate indoor localization. In: 2022 IEEE 12th International Conference on Indoor Positioning and Indoor Navigation (IPIN). pp.~1--8. IEEE (2022)

\bibitem{dl_1}
Qian, W., Lauri, F., Gechter, F.: Supervised and semi-supervised deep probabilistic models for indoor positioning problems. Neurocomputing  \textbf{435},  228--238 (2021)

\bibitem{svm_1}
Rezgui, Y., Pei, L., Chen, X., Wen, F., Han, C.: An efficient normalized rank based svm for room level indoor wifi localization with diverse devices. Mobile Information Systems  \textbf{2017} (2017)

\bibitem{large_labelled_data}
Roy, P., Chowdhury, C.: A survey of machine learning techniques for indoor localization and navigation systems. Journal of Intelligent \& Robotic Systems  \textbf{101}(3), ~63 (2021)

\bibitem{navigation}
Sakpere, W., Adeyeye-Oshin, M., Mlitwa, N.B.: A state-of-the-art survey of indoor positioning and navigation systems and technologies. South African Computer Journal  \textbf{29}(3),  145--197 (2017)

\bibitem{dt_1}
S{\'a}nchez-Rodr{\'\i}guez, D., Hern{\'a}ndez-Morera, P., Quinteiro, J.M., Alonso-Gonz{\'a}lez, I.: A low complexity system based on multiple weighted decision trees for indoor localization. Sensors  \textbf{15}(6),  14809--14829 (2015)

\bibitem{uwb_2}
Segura, M., Mut, V., Sisterna, C.: Ultra wideband indoor navigation system. IET Radar, Sonar \& Navigation  \textbf{6}(5),  402--411 (2012)

\bibitem{helthcare1}
Shum, L.C., Faieghi, R., Borsook, T., Faruk, T., Kassam, S., Nabavi, H., Spasojevic, S., Tung, J., Khan, S.S., Iaboni, A.: Indoor location data for tracking human behaviours: A scoping review. Sensors  \textbf{22}(3), ~1220 (2022)

\bibitem{kNN_better}
Singh, N., Choe, S., Punmiya, R.: Machine learning based indoor localization using wi-fi rssi fingerprints: An overview. IEEE Access  \textbf{9},  127150--127174 (2021)

\bibitem{bluetooth_indoor_1}
Subhan, F., Hasbullah, H., Rozyyev, A., Bakhsh, S.T.: Indoor positioning in bluetooth networks using fingerprinting and lateration approach. In: 2011 International Conference on Information Science and Applications. pp.~1--9. IEEE (2011)

\bibitem{rfid_2}
Ting, S., Kwok, S., Tsang, A.H., Ho, G.T.: The study on using passive rfid tags for indoor positioning. International journal of engineering business management  \textbf{3}(1.),  9--15 (2011)

\bibitem{ujiindoorloc}
Torres-Sospedra, J., Montoliu, R., Mart{\'\i}nez-Us{\'o}, A., Avariento, J.P., Arnau, T.J., Benedito-Bordonau, M., Huerta, J.: Ujiindoorloc: A new multi-building and multi-floor database for wlan fingerprint-based indoor localization problems. In: 2014 international conference on indoor positioning and indoor navigation (IPIN). pp. 261--270. IEEE (2014)

\bibitem{knn_2}
Torteeka, P., Chundi, X.: Indoor positioning based on wi-fi fingerprint technique using fuzzy k-nearest neighbor. In: Proceedings of 2014 11th International Bhurban Conference on Applied Sciences \& Technology (IBCAST) Islamabad, Pakistan, 14th-18th January, 2014. pp. 461--465. IEEE (2014)

\bibitem{wang2019dynamic}
Wang, Y., Sun, Y., Liu, Z., Sarma, S.E., Bronstein, M.M., Solomon, J.M.: Dynamic graph cnn for learning on point clouds. ACM Trans. Graph.  \textbf{38}(5) (oct 2019). \doi{10.1145/3326362}, \url{https://doi.org/10.1145/3326362}

\bibitem{rfid_1}
Xu, H., Wu, M., Li, P., Zhu, F., Wang, R.: An rfid indoor positioning algorithm based on support vector regression. Sensors  \textbf{18}(5), ~1504 (2018)

\bibitem{wifi_1}
Yang, C., Shao, H.R.: Wifi-based indoor positioning. IEEE Communications Magazine  \textbf{53}(3),  150--157 (2015)

\bibitem{dom_adv}
Zhang, M., Fan, Z., Shibasaki, R., Song, X.: Domain adversarial graph convolutional network based on rssi and crowdsensing for indoor localization. arXiv preprint arXiv:2204.05184  (2022)

\end{thebibliography}

\end{document}